\documentclass[prl,twocolumn,showpacs,preprintnumbers,amsmath,amssymb,times]{revtex4}

\usepackage{graphicx}
\usepackage{dcolumn}
\usepackage{bm}
\usepackage{psfrag}
\usepackage{epsfig}
\usepackage{amsmath}
\usepackage{amssymb}
\usepackage{color}
\usepackage{fancyheadings}
\usepackage{mathbbol}

\newcommand{\bra}{\langle}

\newcommand{\ket}{\rangle}
\newcommand{\iu}{\text{i}}
\newcommand{\ee}{\text{e}}

\begin{document}


\title{Effective spin systems in coupled micro-cavities}

\author{Michael J. Hartmann}
\email{m.hartmann@imperial.ac.uk}
\author{Fernando G.S.L. Brand\~ao}
\author{Martin B. Plenio}
\affiliation{Institute for Mathematical Sciences, Imperial College London,
SW7 2PG, United Kingdom}
\affiliation{QOLS, The Blackett Laboratory, Imperial College London, Prince Consort Road,
SW7 2BW, United Kingdom}

\date{\today}

\begin{abstract}
We show that atoms trapped in micro-cavities that interact 
via exchange of virtual photons can model an anisotropic 
Heisenberg spin-$1/2$ chain in an external magnetic field. 
All parameters of the effective Hamiltonian can individually 
be tuned via external lasers.
Since the occupation of excited atomic levels and photonic 
states are strongly suppressed, the effective model is robust 
against decoherence mechanisms, has a long lifetime and its 
implementation is feasible with current experimental technology. 
The model provides a feasible way to create cluster states 
in these devices.
\end{abstract}

\maketitle

%
\paragraph{Introduction:} Interacting two level systems, 
either termed spins or qubits, are of central importance 
in quantum information and condensed matter physics. Two 
or higher dimensional magnetic compounds are believed to 
host some of most interesting condensed matter phenomena, 
such as frustration and high $T_c$ superconductivity \cite{MU03}, 
which are not yet fully understood. In quantum information, 
lattices of interacting spins can be employed to generate 
highly entangled states, such as cluster states which are 
the required resource for one way quantum computation 
\cite{RB01}. While it is a prerequisite for quantum information 
processing, the ability to address individual spins in an 
experimental device can also be very helpful to obtain 
deeper and more detailed insight into condensed matter physics. 

In magnetic compounds where spin lattices appear naturally 
the addressability of individual spins is unfortunately 
extremely hard to achieve because the spatial separation 
between neighboring spins is very small and the timescales 
of interesting processes can be very short.

\begin{figure}[h]
\psfrag{a}{\raisebox{-0.04cm}{$a$}}
\psfrag{b}{\raisebox{-0.04cm}{$b$}}
\psfrag{e}{$e$}
\psfrag{da}{$\delta_a$}
\psfrag{db}{$\delta_b$}
\psfrag{Oa}{\hspace{-0.14cm}$\Omega_a$}
\psfrag{Ob}{$\Omega_b$}
\psfrag{ga}{\hspace{-0.04cm}$g_a$}
\psfrag{gb}{$g_b$}
\psfrag{Da}{\hspace{-0.14cm}$\Delta_a$}
\psfrag{Db}{\hspace{-0.06cm}$\Delta_b$}
\psfrag{wa}{\hspace{-0.1cm}$\omega_a$}
\psfrag{wb}{$\omega_b$}
\psfrag{wab}{$\omega_{ab}$}
\includegraphics[width=7cm]{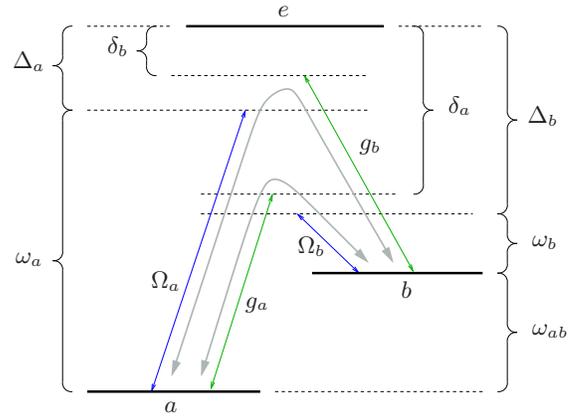}
\caption{\label{xy_levels} Level structure, driving lasers and relevant couplings to the cavity mode to generate effective $\sigma^x \sigma^x$- and $\sigma^y \sigma^y$-couplings for one atom. The cavity mode couples with strengths $g_a$ and $g_b$ to transitions $|a\ket \leftrightarrow |e\ket$ and $|b\ket \leftrightarrow |e\ket$ respectively. One laser with frequency $\omega_a$ couples to transition $|a\ket \leftrightarrow |e\ket$ with Rabi frequency $\Omega_a$ and another laser with frequency $\omega_b$ to $|a\ket \leftrightarrow |e\ket$ with $\Omega_b$. The dominant 2-photon processes are indicated in faint gray arrows.}
\end{figure}

\begin{figure}[h!]
\psfrag{a}{\raisebox{-0.04cm}{$a$}}
\psfrag{b}{\raisebox{-0.04cm}{$b$}}
\psfrag{e}{$e$}
\psfrag{da}{$\tilde{\Delta}_a$}
\psfrag{-db}{$-\tilde{\Delta}_b$}
\psfrag{ka}{\hspace{-0.1cm}$\delta_a$}
\psfrag{kb}{\raisebox{-0.04cm}{\hspace{-0.1cm}$\delta_b$}}
\psfrag{Oa}{\hspace{-0.1cm}$\Omega_a$}
\psfrag{Ob}{\hspace{-0.06cm}$\Omega_b$}
\psfrag{La}{$\Lambda_a$}
\psfrag{Lb}{$\Lambda_b$}
\psfrag{ga}{\hspace{-0.06cm}$g_a$}
\psfrag{gb}{\hspace{-0.1cm}$g_b$}
\psfrag{Da}{\hspace{-0.1cm}$\Delta_a$}
\psfrag{Db}{$\Delta_b$}
\psfrag{n}{$\nu$}
\psfrag{w}{$\omega$}
\psfrag{wb}{$\omega_b$}
\psfrag{wab}{$\omega_{ab}$}
\includegraphics[width=7cm]{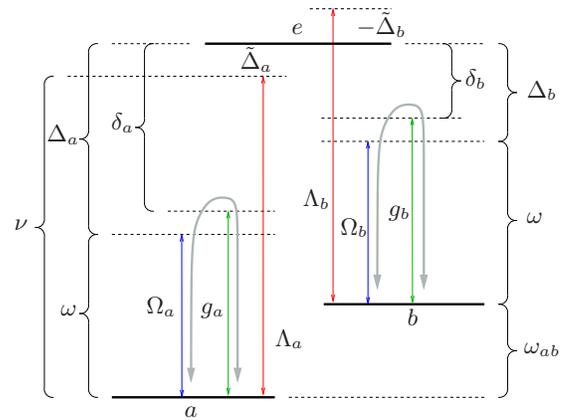}
\caption{\label{zz_levels} Level structure, driving lasers and relevant couplings to the cavity mode to generate effective $\sigma^z \sigma^z$-couplings for one atom. The cavity mode couples with strengths $g_a$
and $g_b$ to transitions $|a\ket \leftrightarrow |e\ket$ and $|b\ket \leftrightarrow |e\ket$ respectively. Two lasers with frequencies $\omega$ and $\nu$ couple with Rabi frequencies $\Omega_a$ respectively $\Lambda_a$ to transition $|a\ket \leftrightarrow |e\ket$ and $\Omega_b$ respectively $\Lambda_b$ to $|b\ket \leftrightarrow |e\ket$. The dominant 2-photon processes are indicated in faint gray arrows.}
\end{figure}

Here we show that effective spin lattices \cite{MW01} can 
be generated with individual atoms in micro-cavities that 
are coupled to each other via the exchange of virtual 
photons. Due to the size and separation of the micro-cavities, 
individual lattice sites can be addressed with optical 
lasers, whereas the cavities can be arranged arbitrarily 
allowing for various lattice geometries. The two spin 
polarizations $|\hspace{-0.1cm}\uparrow \ket$ and 
$|\hspace{-0.1cm}\downarrow \ket$ are thereby represented by two
longlived atomic levels of a $\Lambda$ level-structure
(c.f. figures \ref{xy_levels} and \ref{zz_levels}).
Together with external lasers, the cavity mode that couples 
to these atoms can induce Raman transitions between these two 
long-lived levels. Due to a detuning between laser and cavity 
mode, these transitions can only create virtual photons in 
the cavity mode which mediate an interaction with another 
atom in a neighboring cavity. With appropriately chosen 
detunings, both the excited atomic levels and photon states 
have vanishing occupation and can be eliminated from the 
description. As a result, the dynamics is confined to only 
two states per atom, the long-lived levels, and can be 
described by a spin-1/2 Hamiltonian.
Due to the small occupation of photon states and excited 
atomic levels, spontaneous emission and cavity decay are 
strongly suppressed. All these results are verified by 
detailed numerics.
A realization of the scheme thus requires cavities that 
operate in a strong coupling regime with a high cooperativity 
factor and an atom photon coupling that exceeds cavity decay.
Such regimes have now been achieved in several devices 
\cite{ADW+06,BHA+05,THE+05}, making a realization of the 
presented scheme feasible with current technology. We begin 
by showing how to engineer effective $\sigma^x \sigma^x$, 
$\sigma^y \sigma^y$ and $\sigma^z \sigma^z$ interactions 
as well as the effective magnetic field $\sigma_z$ and
then explain how to generate the full anisotropic Heisenberg 
model. We verify the validity of all approximations by 
comparison with the exact dynamics of the full atom-cavity 
model and also apply the model to the generation of cluster 
states. Finally we discuss the feasibility of our model for
realistic experimental parameters.

\paragraph{XX and YY interactions:}

We consider an array of cavities that are coupled via exchange 
of photons with one 3-level atom in each cavity (figure \ref{xy_levels}).
Two long lived levels, $|a\ket$ and $|b\ket$, represent the 
two spin states. The cavity mode couples to the transitions 
$|a\ket \leftrightarrow |e\ket$ and $|b\ket \leftrightarrow |e\ket$, 
where $|e\ket$ is the excited state of the atom. Furthermore, 
two driving lasers couple to the transitions 
$|a\ket \leftrightarrow |e\ket$ respectively $|b\ket \leftrightarrow |e\ket$.
The Hamiltonian of the atoms reads 
$H_A = \sum_{j=1}^N \omega_e |e_j\ket \bra e_j| + \omega_{ab} |b_j\ket \bra b_j|$, 
where the index $j$ counts the cavities, $\omega_e$ is the 
energy of the excited level and $\omega_{ab}$ the energy of 
level $|b\ket$. The energy of level $|a\ket$ is set to zero 
and we use $\hbar = 1$. The Hamiltonian that describes the 
photons in the cavity modes is 
$H_C = \omega_C \sum_{j=1}^N a_j^{\dagger} a_j + J_C \sum_{j=1}^N \left( a_j^{\dagger} a_{j+1} + a_j a_{j+1}^{\dagger} \right)$, 
where $a_j^{\dagger}$ creates a photon in cavity $j$, 
$\omega_C$ is the energy of the photons and $J_C$ the 
tunneling rate of photons between neighboring cavities 
\cite{HBP06}. For convenience we assume periodic boundary 
conditions, where $H_C$ can be diagonalized via the Fourier 
transform 
$a_k = \frac{1}{\sqrt{N}} \sum_{j=1}^N \ee^{\iu k j} a_j \: ; \: k = \frac{2 \pi l}{N} \: ; \: - \frac{N}{2} \le l \le \frac{N}{2} \quad (N \, \text{odd})$ to give $H_C = \sum_{k} \omega_k a_k^{\dagger} a_k$ with $\omega_k = \omega_C + 2 J_C \cos(k)$. 
Finally the interaction between the atoms and the photons 
as well as the driving by the lasers are described by 
$H_{AC} = \sum_{j=1}^N \left[ \left(\frac{1}{2} \Omega_a \ee^{-\iu \omega_a t} + g_a a_j \right) |e_j\ket \bra a_j| + \text{h.c.} \right] +\linebreak
+ [ a \leftrightarrow b ]$.
Here $g_a$ and $g_b$ are the couplings of the respective transitions to the cavity mode, $\Omega_a$ is the Rabi frequency of one laser with frequency $\omega_a$ and  $\Omega_b$
the Rabi frequency of a second laser with frequency $\omega_b$ \cite{SM02}. The complete Hamiltonian is then given by $H = H_A + H_C + H_{AC}$ .

We now switch to an interaction picture with respect to $H_0 = H_A + H_C - \delta_1 \, \sum_{j=1}^N |b_j\ket \bra b_j|$, where $\delta_1 = \omega_{ab} - (\omega_a - \omega_b)/2$, and adiabatically eliminate the excited atom levels $|e_j\ket$ and the photons \cite{J00}. We consider terms up to 2nd order in the effective Hamiltonian and drop fast oscillating terms. For this approach the detunings $\Delta_a \equiv \omega_e - \omega_a$, $\Delta_b \equiv \omega_e - \omega_b - (\omega_{ab} - \delta_1)$, $\delta_a^k \equiv \omega_e - \omega_k$ and $\delta_b^k \equiv \omega_e - \omega_k - (\omega_{ab} - \delta_1)$ have to be large compared to the couplings $\Omega_a, \Omega_b, g_a$ and $g_b$, i.e. $|\Delta_a|, |\Delta_b|, |\delta_a^k|, |\delta_b^k| \gg |\Omega_a|, |\Omega_b|, |g_a|, |g_b|$ (for all $k$). Furthermore, the parameters must be such that the dominant Raman transitions between levels $a$ and $b$ are those that involve one laser photon and one cavity photon each (c.f. figure \ref{xy_levels}). To avoid excitations of real photons via these transitions, we furthermore require 
$\left| \Delta_a - \delta_b^k \right|, \left| \Delta_b - \delta_a^k \right| \gg \left| \frac{\Omega_a g_b}{2 \Delta_a} \right|, \left| \frac{\Omega_b g_a}{2 \Delta_b} \right| $ (for all $k$).
 
Hence whenever the atom emits or absorbs a virtual photon 
into or from the cavity mode, it does a transition from 
level $|a\ket$ to $|b\ket$ or vice versa. If one atom emits 
a virtual photon in such a process that is absorbed by a 
neighboring atom, which then also does a transition between 
$|a\ket$ to $|b\ket$, an effective spin-spin interaction 
has happened. Dropping irrelevant constants, the resulting 
effective Hamiltonian reads
$$ H_{\text{xy}} = \sum_{j=1}^N B \sigma_j^z
+ \left( J_1 \sigma_j^+ \sigma_{j+1}^- + J_2 \sigma_j^- \sigma_{j+1}^- + \text{h.c.} \right),$$
where $\sigma_j^z = |b_j\ket \bra b_j| - |a_j\ket \bra a_j|$ 
and $\sigma_j^+ = |b_j\ket \bra a_j|$. The parameters $B$, 
$J_1$ and $J_2$ are given to second order by \cite{paramXY}.
If $J_2^{\star} = J_2$, this Hamiltonian reduces to the XY model,
\begin{equation} \label{HXYb}
H_{\text{xy}} = \sum_{j=1}^N B \sigma_j^z + J_x \sigma_j^x 
\sigma_{j+1}^x + J_y \sigma_j^y \sigma_{j+1}^y \, ,
\end{equation}
with $J_x = (J_1 + J_2)/2$ and $J_y = (J_1 - J_2)/2$.

For $\Omega_a = \pm (\Delta_a g_a / \Delta_b g_b) \Omega_b$ 
with $\Omega_a$ and $\Omega_b$ real, the interaction is either 
purely $\sigma^x \sigma^x$ ($+$) or purely $\sigma^y \sigma^y$ 
($-$) and the Hamiltonian (\ref{HXYb}) becomes the Ising model 
in a transverse field, whereas the isotropic XY model 
($J_x = J_y$) \cite{DDL03} is obtained for either 
$\Omega_a \rightarrow 0$ or $\Omega_b \rightarrow 0$. The 
effective magnetic field $B$ in turn can, independently of 
$J_x$ and $J_y$, be tuned to assume any value between 
$|B| \gg |J_x|, |J_y|$ and $|B| \ll |J_x|, |J_y|$ by varying 
$\delta_1$. Thus we will be able to drive the system through
a quantum phase transition. Now we proceed to show how to
engineer effective ZZ interactions.

\paragraph{ZZ interactions:}

To obtain an effective $\sigma^z \sigma^z$ interaction, we again use the same atomic level configuration but now only one laser with frequency $\omega$ mediates atom-atom coupling via virtual photons. A second laser with frequency $\nu$ is used to tune the effective magnetic field via a Stark shift. The atoms together with their couplings to cavity mode and lasers are shown in figure \ref{zz_levels}.
The Hamiltonians $H_A$ of the atoms and $H_C$ of the cavity modes thus have the same form as above, whereas $H_{AC}$ now reads:\linebreak
$H_{AC} = \sum_{j=1}^N \left[ \left(\frac{\Omega_a}{2} \ee^{-\iu \omega t} +
\frac{\Lambda_a}{2} \ee^{-\iu \nu t} + g_a a_j \right) |e_j\ket \bra a_j| + \text{h.c.} \right] \linebreak
+  \left[ a \leftrightarrow b \right]$ .
Here, $\Omega_a$ and $\Omega_b$ are the Rabi frequencies of the driving laser with frequency $\omega$ on transitions $|a\ket \rightarrow |e\ket$ and $|b\ket \rightarrow |e\ket$, whereas $\Lambda_a$ and $\Lambda_b$ are the Rabi frequencies of the driving laser with frequency $\nu$ on transitions $|a\ket \rightarrow |e\ket$ and $|b\ket \rightarrow |e\ket$.

We switch to an interaction picture with respect to $H_0 = H_A + H_C$ and adiabatically eliminate the excited atom levels $|e_j\ket$ and the photons \cite{J00}. Again, the detunings $\Delta_a \equiv \omega_e - \omega$, $\Delta_b \equiv \omega_e - \omega - \omega_{ab}$, $\tilde{\Delta}_a \equiv \omega_e - \nu$, $\tilde{\Delta}_b \equiv \omega_e - \nu - \omega_{ab}$, $\delta_a^k \equiv \omega_e - \omega_k$ and $\delta_b^k \equiv \omega_e - \omega_k - \omega_{ab}$ have to be large compared to the couplings $\Omega_a, \Omega_b, \Lambda_a, \Lambda_b, g_a$ and $g_b$, i.e. $|\Delta_a|, |\Delta_b|, |\delta_a^k|, |\delta_b^k| \gg |\Omega_a|, |\Omega_b|, |g_a|, |g_b|$ and $|\tilde{\Delta}_a|, |\tilde{\Delta}_b| \gg |\Lambda_a|, |\Lambda_b|$ (for all $k$), whereas now Raman transitions between levels $a$ and $b$ should be suppressed. Hence parameters must be such that the dominant 2-photon processes are those that involve one laser photon and one cavity photon each but where the atom does no transition between levels $a$ and $b$ (c.f. figure \ref{zz_levels}). To avoid excitations of real photons in these processes, we thus require 
$\left| \Delta_a - \delta_a^k \right|, \left| \Delta_b - \delta_b^k \right| \gg \left| \frac{\Omega_a g_a}{2 \Delta_a} \right|, \left| \frac{\Omega_b g_b}{2 \Delta_b} \right| $ (for all $k$).

Whenever two atoms exchange a virtual photon in this scheme, none of them does a transition between $|a\ket$ and $|b\ket$. Moreover both atoms experience a Stark shift that depends on the state of the partner atom. This conditional Stark shifts play the role of an effective $\sigma^z \sigma^z$-interaction.
Dropping irrelevant constants, the resulting effective Hamiltonian reads:
\begin{equation} \label{HZZ}
H_{\text{zz}} = \sum_{j=1}^N \left( \tilde{B} \sigma_j^z + J_z \sigma_j^z \sigma_{j+1}^z \right) \, ,
\end{equation}
where the parameters $\tilde{B}$ and $J_z$ are given to second order by \cite{paramZZ}.
Here again, the interaction $J_z$ and the field $\tilde{B}$ can be tuned independently, either by varying
$\Omega_a$ and $\Omega_b$ for $J_z$ or by varying $\Lambda_a$ and $\Lambda_b$ for $\tilde{B}$. In particular,
$|\Lambda_a|^2$ and $|\Lambda_b|^2$ can for all values of $\Omega_a$ and $\Omega_b$ be chosen such that either $J_z \ll \tilde{B}$ or $J_z \gg \tilde{B}$. 

\paragraph{The complete effective model:} 

Making use of the Suzuki-Trotter formula, the two 
Hamiltonians (\ref{HXYb}) and (\ref{HZZ}) can now be combined 
to one effective Hamiltonian. To this end, the lasers that 
generate the Hamiltonian (\ref{HXYb}) are turned on for a 
short time interval $dt$ ($||H_{\text{xy}}|| \cdot dt \ll 1$) 
followed by another time interval $dt$ ($||H_{\text{zz}}|| \cdot dt \ll 1$) with the lasers that generate the Hamiltonian (\ref{HZZ}) 
turned on. This sequence is repeated until the total time 
range to be simulated is covered. The effective Hamiltonian 
simulated by this procedure is 
$H_{\text{spin}} = H_{\text{xy}} + H_{\text{zz}}$ or
\begin{equation} \label{Hspin}
H_{\text{spin}} = \sum_{j=1}^N \left( B_{\text{tot}} \sigma_j^z + \sum_{\alpha = x,y,z} J_{\alpha} \sigma_j^{\alpha} \sigma_{j+1}^{\alpha} \right) \, ,
\end{equation}
where $B_{\text{tot}} = B + \tilde{B}$. The time interval 
$dt$ should thereby be chosen such that 
$\Omega^{-1}, g^{-1} \ll  dt_1 , dt_2 \ll J_x^{-1} , J_y^{-1} , J_z^{-1} , B^{-1}$ and $\tilde{B}^{-1}$, so that the Trotter sequence 
concatenates the effective Hamiltonians $H_{XY}$ and $H_{ZZ}$. 
The procedure can be generalized to higher order Trotter formulae
or by turning on the sets of lasers for time intervals of different 
length.

\paragraph{Numerical tests:} 

To confirm the validity of our approximations, we numerically 
simulate the dynamics generated by the full Hamiltonian $H$ 
and compare it to the dynamics generated by the effective model 
(\ref{Hspin}).

\begin{figure}[h!]
\psfrag{t}{\raisebox{-0.4cm}{\scriptsize \hspace{-0.6cm} $t \: \text{in} \: 10^{-6} \: \text{s}$}}
\psfrag{pa}{\raisebox{0.4cm}{\scriptsize \hspace{-0.6cm} $p(a_1), \: p( \downarrow_1 )$}}
\psfrag{A}{\bf \hspace{-1.6cm} b}
\psfrag{B}{\bf \hspace{-1.6cm} a}
\psfrag{0a}{\raisebox{-0.12cm}{\scriptsize $0$}}
\psfrag{5a}{\raisebox{-0.12cm}{\scriptsize $5$}}
\psfrag{10a}{\raisebox{-0.12cm}{\scriptsize $10$}}
\psfrag{15a}{\raisebox{-0.12cm}{\scriptsize $15$}}
\psfrag{20a}{\raisebox{-0.12cm}{\scriptsize $20$}}
\psfrag{25a}{\raisebox{-0.12cm}{\scriptsize $25$}}
\psfrag{30a}{\raisebox{-0.12cm}{\scriptsize $30$}}
\psfrag{40a}{\raisebox{-0.12cm}{\scriptsize $40$}}
\psfrag{50a}{\raisebox{-0.12cm}{\scriptsize $50$}}
\psfrag{60a}{\raisebox{-0.12cm}{\scriptsize $60$}}
\psfrag{0}{\hspace{-0.3cm} \scriptsize $0$}
\psfrag{0.1}{\hspace{-0.4cm} \scriptsize $ $}
\psfrag{0.2}{\hspace{-0.4cm} \scriptsize $0.2$}
\psfrag{0.3}{\hspace{-0.4cm} \scriptsize $ $}
\psfrag{0.4}{\hspace{-0.4cm} \scriptsize $0.4$}
\psfrag{0.5}{\hspace{-0.4cm} \scriptsize $ $}
\psfrag{0.6}{\hspace{-0.4cm} \scriptsize $0.6$}
\psfrag{0.7}{\hspace{-0.4cm} \scriptsize $ $}
\psfrag{0.8}{\hspace{-0.4cm} \scriptsize $0.8$}
\psfrag{0.9}{\hspace{-0.4cm} \scriptsize $ $}
\psfrag{1}{\hspace{-0.3cm} \scriptsize $1$}
\hspace{0.2cm}
\includegraphics[width=4cm]{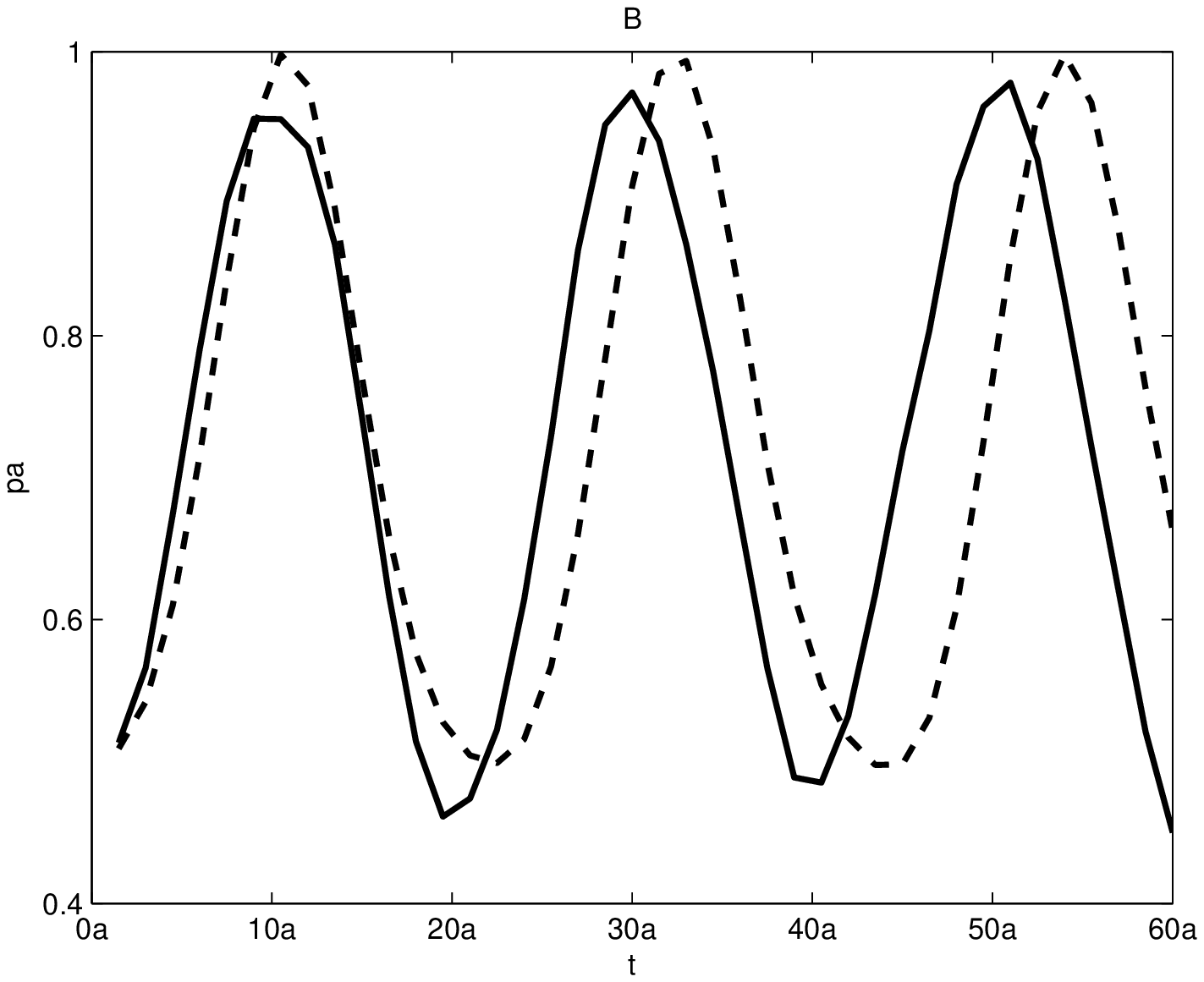}
\hspace{0.1cm}
\includegraphics[width=4cm]{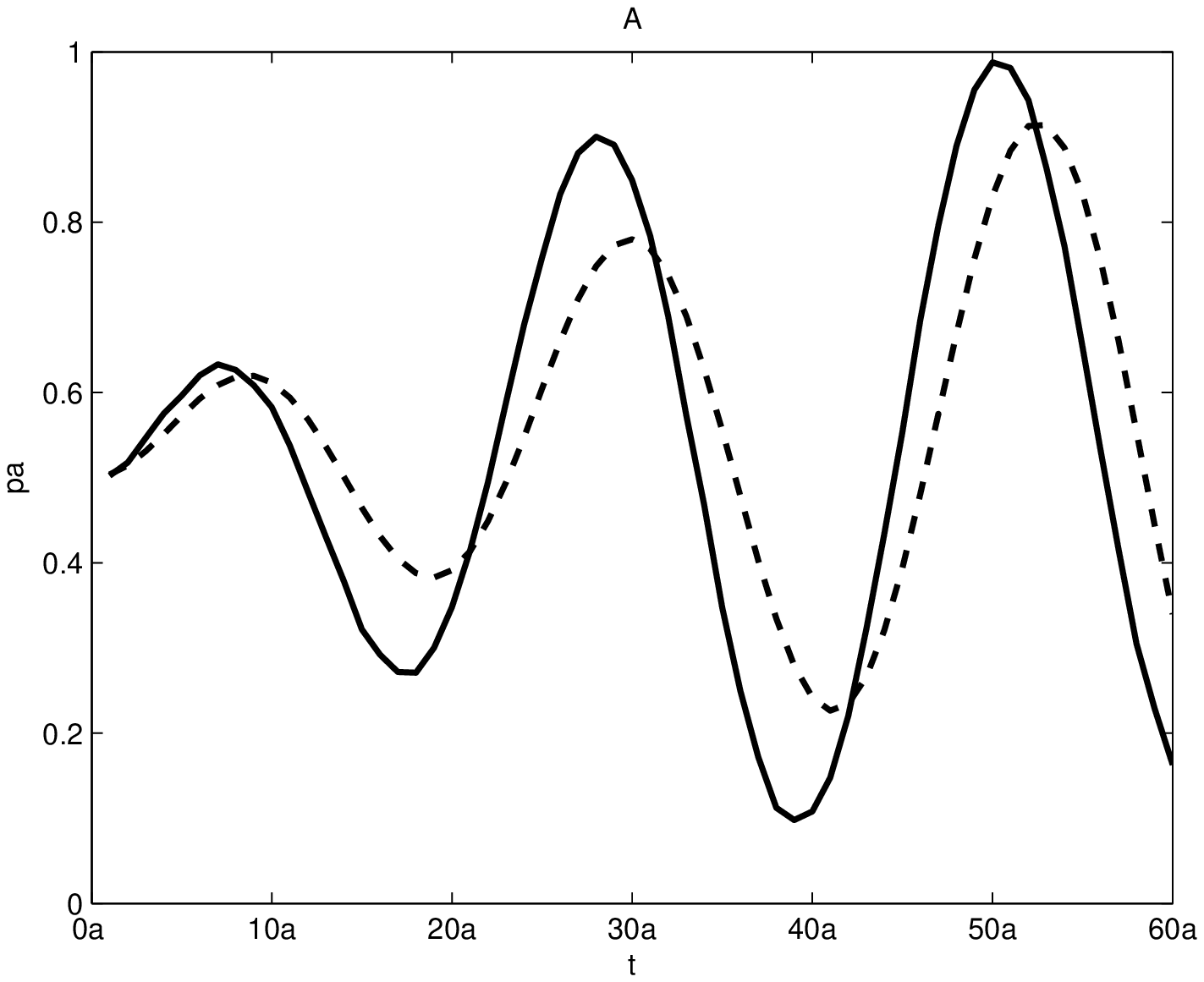}
\caption{\label{run2} The occupation probability $p(a_1)$ of state $|a_1\ket$ (solid line) and the
probability $p( \downarrow_1 )$ of spin 1 to point down (dashed line) for the parameters
$\omega_e = 10^6$GHz, $\omega_{ab} = 30$GHz, 
$\Delta_a = 30$GHz, $\Delta_b = 60$GHz, $\omega_C = \omega_e - \Delta_b + 2$GHz, $\tilde{\Delta}_a = 15$GHz, $\Omega_a = \Omega_b = 2$GHz, $\Lambda_a = \Lambda_b = 0.71$GHz, $g_a = g_b = 1$GHz, $J_C = 0.2$GHz and
$\delta_1 = -0.0165$GHz (plot {\bf a}) respectively $\delta_1 = -0.0168$GHz (plot {\bf b}). Both, the occupation of the excited atomic states $\bra |e_j\ket \bra e_j | \ket$ and the photon number $\bra a^{\dagger} a \ket$ are always smaller than 0.03.}
\end{figure}

As an example we consider two atoms in two cavities, initially 
in the state $\frac{1}{\sqrt{2}} (|a_1\ket + |b_1\ket) \otimes |a_2\ket$, and calculate the occupation probability $p(a_1)$ of the state 
$|a_1\ket$ which corresponds to the probability of spin 1 to 
point down, $p( \downarrow_1 )$. Figure \ref{run2}{\bf a} shows 
$p(a_1)$ and $p( \downarrow_1 )$ for an effective Hamiltonian 
(\ref{Hspin}) with $B_{\text{tot}} = 0.135$MHz, $J_x = 0.065$MHz, 
$J_y = 0.007$MHz and $J_z = 0.004$MHz and hence 
$|B_{\text{tot}}| > |J_x|$, whereas figure \ref{run2}{\bf b} 
shows $p(a_1)$ and $p( \downarrow_1 )$ for an effective Hamiltonian 
(\ref{Hspin}) with $B_{\text{tot}} = -0.025$MHz, $J_x = 0.065$MHz, 
$J_y = 0.007$MHz and $J_z = 0.004$MHz and hence 
$|B_{\text{tot}}| < |J_x|$ \cite{HRP06}.

Discrepancies between numerical results for the full and 
the effective model are due to higher order terms for the 
parameters \cite{paramXY,paramZZ}, which lead to relative 
corrections of up to 10\% in the considered cases. Let us 
stress here that despite this lack of accuracy of the 
approximations \cite{paramXY,paramZZ}, the effective model 
is indeed a spin-1/2 Hamiltonian as occupations of excited 
atomic and photon states are negligible.

\paragraph{Cluster state generation:} 

\begin{figure}[h!]
\psfrag{A}{\bf \hspace{-1.6cm} a}
\psfrag{B}{\bf \hspace{-1.6cm} b}
\psfrag{t}{\raisebox{-0.4cm}{\scriptsize \hspace{-0.6cm} $t \: \text{in} \: 10^{-6} \: \text{s}$}}
\psfrag{pu}{\raisebox{0.4cm}{\scriptsize \hspace{-0.2cm} $P_{\text{s}}$}}
\psfrag{vN}{\raisebox{0.4cm}{\scriptsize \hspace{-0.6cm} $E_{\text{vN}} / \ln 2$}}
\psfrag{0a}{\raisebox{-0.12cm}{\scriptsize $0$}}
\psfrag{5a}{\raisebox{-0.12cm}{\scriptsize $5$}}
\psfrag{10a}{\raisebox{-0.12cm}{\scriptsize $10$}}
\psfrag{15a}{\raisebox{-0.12cm}{\scriptsize $15$}}
\psfrag{20a}{\raisebox{-0.12cm}{\scriptsize $20$}}
\psfrag{25a}{\raisebox{-0.12cm}{\scriptsize $25$}}
\psfrag{0}{\hspace{-0.4cm} \scriptsize $0$}
\psfrag{0.2}{\hspace{-0.4cm} \scriptsize $0.2$}
\psfrag{0.4}{\hspace{-0.4cm} \scriptsize $0.4$}
\psfrag{0.6}{\hspace{-0.4cm} \scriptsize $0.6$}
\psfrag{0.8}{\hspace{-0.4cm} \scriptsize $0.8$}
\psfrag{1}{\hspace{-0.4cm} \scriptsize $1$}
\psfrag{0.92}{\hspace{-0.5cm} \scriptsize $0.92$}
\psfrag{0.94}{\hspace{-0.5cm} \scriptsize $0.94$}
\psfrag{0.96}{\hspace{-0.5cm} \scriptsize $0.96$}
\psfrag{0.98}{\hspace{-0.5cm} \scriptsize $0.98$}
\hspace{0.2cm}
\includegraphics[width=4cm]{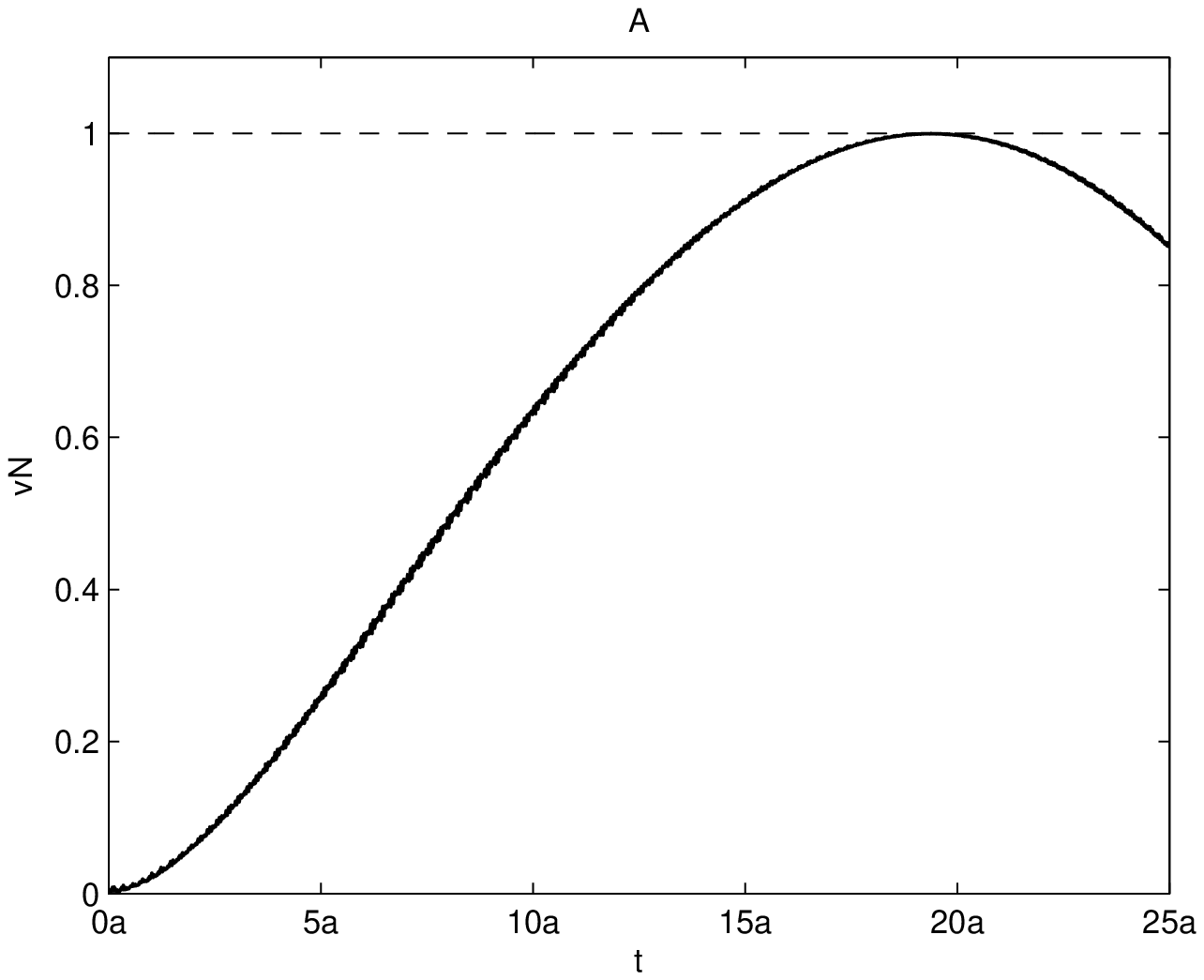}
\hspace{0.1cm}
\includegraphics[width=4cm]{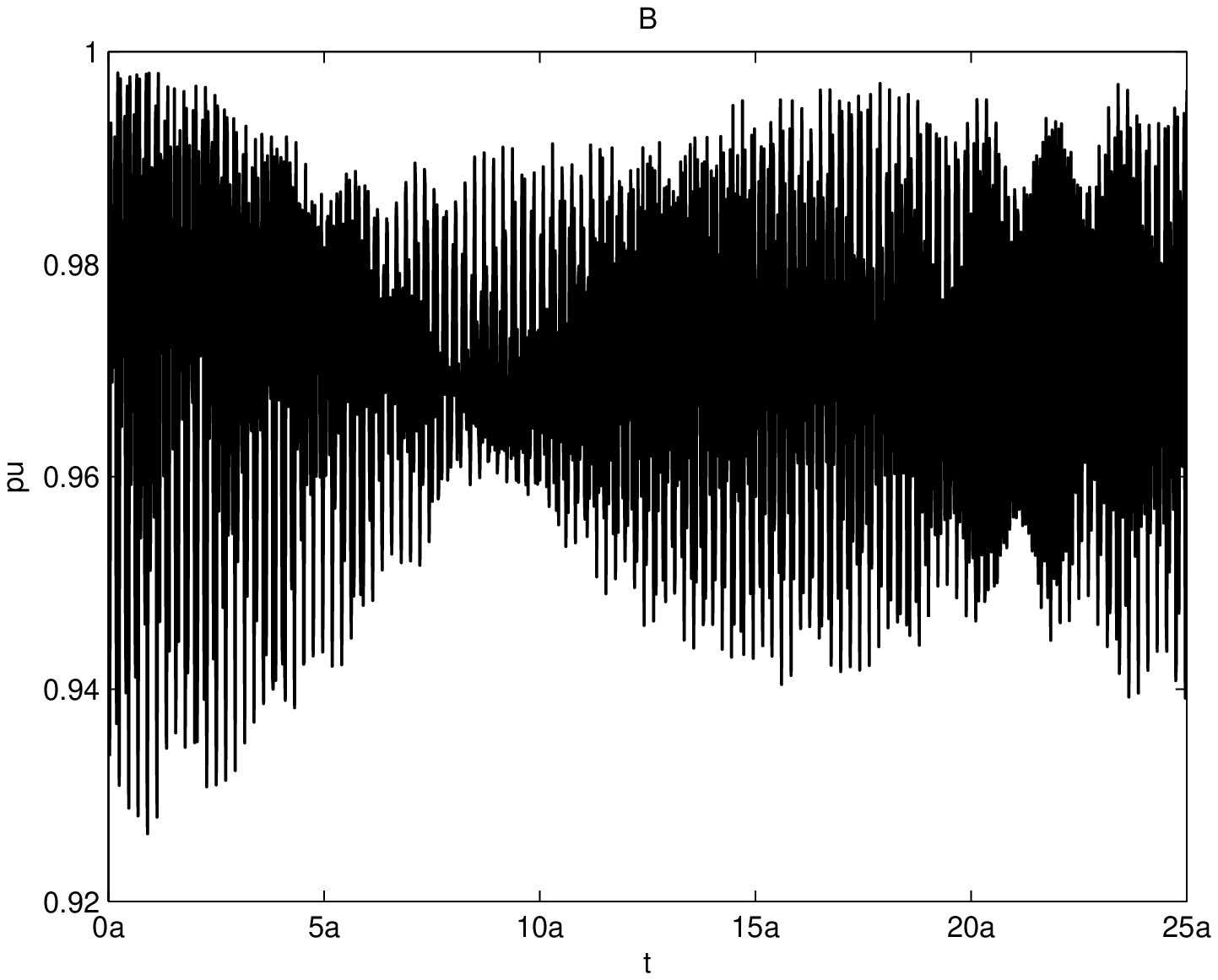}
\caption{\label{cluster} {\bf a} The von Neumann entropy $E_{\text{vN}}$ of the reduced density matrix of 1 effective spin in multiples of $\ln 2$ and {\bf b} the purity of the reduced state of the effective spin model for 2 cavities where $J_z = 0.042$MHz.}
\end{figure}

The Hamiltonian (\ref{HZZ}) can be used to generate cluster 
states \cite{RB01,AK07}. To this end, all atoms are initialized 
in the states $(|a_j\ket + |b_j\ket)/\sqrt{2}$, which can 
be done via a STIRAP process \cite{FIM05}, and then evolved 
under the Hamiltonian (\ref{HZZ}) for $t = \pi / 4 J_z$.
Figure \ref{cluster} shows the von Neumann entropy of the 
reduced density matrix of one effective spin $E_{\text{vN}}$ 
and the purity of the reduced density matrix of the effective 
spin chain $P_{\text{s}}$ for a full two cavity model. Since 
$E_{\text{vN}} \approx log_2 2$ for $t \approx 19 \mu$s while 
the state of the effectve spin model remains highly pure 
($P_{\text{s}} = tr[\rho^2] > 0.95$) the degree of entanglement
will be very close to maximal, see e.g. \cite{Audenaert P 06}. 
Thus the levels  $|a_j\ket$ and $|b_j\ket$ have indeed been 
driven into a state which is, up to local unitary rotations, 
very close to a two-qubit cluster state a.k.a. singlet states.

\paragraph{Experimental implementation:} 

For an experimental implementation, the parameters of the 
effective Hamiltonian, $J_x$, $J_y$, $J_z$, $B$ and $\tilde{B}$ 
have to bemuch larger than rates for decay mechanisms via 
the photons or the excited states $| e_j \ket$. 

With the definitions $\Omega = \text{max}(\Omega_a, \Omega_b)$, $g = \text{max}(g_a, g_b)$, $\Delta = \text{min}(\Delta_a, \Delta_b)$, the occupation of the excited levels $| e_j \ket$ can be estimated to be $\langle | e_j \ket \bra e_j | \rangle \approx |\Omega / 2 \Delta |^2$, whereas the photon number is $n_p \approx |(\Omega g / 2 \Delta) \gamma_1 |^2$ and the couplings $J_x$, $J_y$ and $J_z$ are approximately $|(\Omega g / 2 \Delta)|^2 \gamma_2$. Spontaneous emission from the levels $| e_j \ket$ at a rate $\Gamma_E$ and cavity decay of photons at a rate $\Gamma_C$ thus lead to decay rates $\Gamma_1 = |\Omega / 2 \Delta |^2 \Gamma_E$ and $\Gamma_2 = |(\Omega g / 2 \Delta) \gamma_1 |^2 \Gamma_C$ for the effective model.

Hence, we require $\Gamma_1 \ll |(\Omega g / 2 \Delta)|^2 \gamma_2$ and $\Gamma_2 \ll |(\Omega g / 2 \Delta)|^2 \gamma_2$ which implies $\Gamma_E \ll J_C \, g^2 / \delta^2$ and $\Gamma_C \ll J_C$
($J_C < \delta/2$), where, $\delta =  |(\omega_a + \omega_b)/2 - \omega_C|$ for the XX and YY interactions and $\delta =  |\omega - \omega_C|$ for the ZZ interactions and we have approximated
$|\gamma_1| \approx \delta^{-1}$ and $|\gamma_2| \approx J_C \delta^{-2}$.
Since photons should be more likely to tunnel to the next cavity than decay into free space,
$\Gamma_C \ll J_C$ should hold in most cases. For $\Gamma_E \ll J_C g^2 / \delta^2$,
to hold, cavities with a high ratio $g / \Gamma_E$ are favorable. Since $\delta > 2 J_C$, the two requirements together imply that the cavities should have a high cooperativity factor.

This regime can be achieved in micro-cavities, which have a small volume and thus a high $g$. Suitable candidates for the present proposal are for example photonic band gap cavities \cite{BHA+05} which can either couple to atoms or quantum dots. Here, cooperativity factors of $g^2/ 2 \Gamma_C \Gamma_E \sim 10$ and values of $g / \Gamma_E \sim 100$ have been realized and $g^2/ 2 \Gamma_C \Gamma_E \sim 10^5$ respectively  $g / \Gamma_E \sim 10^5$ are predicted to be achievable \cite{SKV+05}. Further promising devices are micro-cavities on a gold coated silicon chip that couple to single trapped atoms, where $g^2/ 2 \Gamma_C \Gamma_E \sim 40$ and  $g / \Gamma_E \sim 50$ have been achieved \cite{THE+05}.
Both are fabricated in large arrays and couple via the overlap 
of their evanescent fields or optical fibers that transfer 
photons from one cavity to another.

\paragraph{Summary:} 

We have shown that single atoms in interacting cavities 
that are operated in a strong coupling regime can form a 
Heisenberg spin-1/2 Hamiltonian. All parameters of the 
effective Hamiltonian can be tuned individually, making 
the device a universal simulator for this model. When 
operated in a two dimensional array of cavities the 
device is thus able to simulate spin lattices which are 
not trackable with numerics on classical computers. Furthermore,
this system can be used to generate cluster states on 
such lattices. Together with the possibility to measure 
individual lattice sites it thus provides the two key 
requirements for one way quantum computation. This 
demonstrates the versatility of the present set-up for
the control and manipulation of quantum systems in
parameter ranges that are experimentally accessible.
%
%

\end{document}